\documentclass[aps,pre,10pt,showpacs,floatfix,onecolumn,nofootinbib,a4paper]{revtex4-2}

\usepackage{amsmath}
\usepackage{amssymb}
\usepackage{amsfonts}
\usepackage{graphicx,color}
\usepackage{bm}
\usepackage{mathrsfs}
\usepackage{verbatim}
\usepackage{url}
\usepackage{caption}
\usepackage{subcaption}

\renewcommand{\tensor}[1]{\mathbf{#1}}
\newcommand{\diver}{\operatorname{div}}
\newcommand{\curl}{\operatorname{curl}}
\newcommand{\tr}{\operatorname{tr}}

\newcommand{\sgn}{\operatorname{sgn}}
\newcommand{\n}{\bm{n}}
\newcommand{\x}{\bm{x}}
\newcommand{\e}{\bm{e}}

\newcommand{\bend}{\bm{b}}
\newcommand{\bendhat}{\widehat{\bend}}
\newcommand{\trans}{^\mathsf{T}}
\newcommand{\R}{\mathbf{R}}
\newcommand{\W}{\mathbf{W}}
\newcommand{\oct}{\bm{\mathbf{A}}}
\newcommand{\id}{\mathbf{I}}

\newcommand{\bsplay}{\mathbf{D}}
\newcommand{\Proj}{\mathbf{P}}
\newcommand{\gradn}{\nabla\n}
\newcommand{\spin}{\bm{\omega}}
\newcommand{\Spin}{\mathbf{\Omega}}
\newcommand{\con}{\bm{c}}
\newcommand{\cond}{\bm{d}}
\newcommand{\zero}{\bm{0}}
\newcommand{\dframe}{(\n_1,\n_2,\n)}

\newcommand{\traj}{\mathscr{C}}
\newcommand{\body}{\mathscr{B}}
\newcommand{\boundary}{\partial\body}

\newcommand{\real}{\mathbb{R}}
\newcommand{\complex}{\mathbb{C}}
\newcommand{\sphere}{\mathbb{S}}
\newcommand{\NTB}{$N_\mathrm{TB}$}
\newcommand{\FTB}{F_\mathrm{TB}}
\newcommand{\FF}{F_\mathrm{F}}
\newcommand{\fTB}{f_\mathrm{TB}}
\newcommand{\measures}{(S,T,\bend,\bsplay)}

\newlength{\irrl}
\newlength{\irrw}

\newcommand{\hess}{\nabla^2}
\newcommand{\vv}{\bm{v}}

\begin{document}

\title{Uniform distortions and generalized elasticity of liquid crystals}
\author{Epifanio G. Virga}
\email{eg.virga@unipv.it}
\affiliation{Dipartimento di Matematica, Universit\`a di Pavia, Via Ferrata 5, 27100 Pavia, Italy}


\begin{abstract}
Ordinary nematic liquid crystals are characterized by having a uniform director field as ground state. In such a state, the director is the same everywhere and no distortion is to be seen at all. We give a definition of uniform distortion which makes precise the intuitive notion of seeing everywhere the same director landscape. We characterize all such distortions and prove that they fall into two families, each described by two scalar parameters. Uniform distortions exhaust R. Meyer's \emph{heliconical} structures, which, as it has recently been recognized, include the ground state of twist-bend nematics. The generalized elasticity of these new phases is treated with a simple free-energy density, which can be minimized by both uniform and non-uniform distortions, the latter injecting a germ of elastic frustration. 
\end{abstract}

\maketitle

\section{Introduction}\label{sec:intro}
More often than not, a new fresh look into an established theory reveals unexpected scenarios, which a sedimented knowledge had prevented from seeing. This is certainly the case of a paper by Selinger \cite{selinger:interpretation} on the reinterpretation of Frank's elastic theory for liquid crystals \cite{frank:theory}.

A unit vector, the nematic director $\n$, is the sole player of this theory, which hinges on a free-energy density expressed by the most general quadratic form in $\gradn$ invariant under change of observer and enjoying the nematic symmetry (which reverses the sign of $\n$). Frank's formula is the following
\begin{equation}\label{eq:Frank_energy_density}
F_\mathrm{F}=\frac12K_{11}(\diver\n)^2+\frac12K_{22}(\n\cdot\curl\n)^2+\frac12K_{33}|\n\times\curl\n|^2+K_{24}(\tr(\gradn)^2-(\diver\n)^2),
\end{equation}
where $K_{11}$, $K_{22}$, $K_{33}$, $K_{24}$ are Frank's elastic constants. As  remarked by Ericksen \cite{ericksen:nilpotent}, the $K_{24}$ term, which is called the \emph{saddle-splay} energy, has a special status, which also justifies the way invariants are grouped in \eqref{eq:Frank_energy_density}. That term is a \emph{null Lagrangian}, which can be integrated over the domain $\body$ occupied by the material, contributing nothing to the total energy whenever $\n$ is appropriately prescribed on $\boundary$.\footnote{For example, when $\n$ is strongly anchored  on $\boundary$, see also \cite[Chap.\,3]{virga:variational}.} The other contributions to $\FF$ are genuine bulk terms; they are the \emph{splay}, \emph{twist}, and \emph{bend} energies, respectively.

The starting point of \cite{selinger:interpretation} is a decomposition of $\gradn$ already found in \cite{machon:umbilic}, which however had a different pursuit. Letting the scalar $S:=\diver\n$  be the \emph{splay}, the pseudo-scalar $T:=\n\cdot\curl\n$ be the \emph{twist}, and the vector $\bend:=\n\times\curl\n$ be the \emph{bend}, and denoting by $\W(\n)$ the shew-symmetric tensor\footnote{$\W(\n)$ acts on any vector $\vv$ as a cross product, $\W(\n)\vv=\n\times\vv$.} associated with $\n$ and by $\Proj(\n):=\id-\n\otimes\n$ the projector on the plane orthogonal to $\n$, one proves the identity \cite{selinger:interpretation}
\begin{equation}\label{eq:decomposition_gradient}
\gradn=-\bend\otimes\n+\frac12T\W(\n)+\frac12S\Proj(\n)+\bsplay,
\end{equation}
where $\bsplay$ is a symmetric tensor such that $\bsplay\n=\zero$ and $\tr\bsplay=0$.\footnote{What here is $\bsplay$ was $\bm{\Delta}$ in \cite{machon:umbilic} and \cite{selinger:interpretation}. I cannot help disliking to mix Latin and Greek alphabets in \eqref{eq:decomposition_gradient}.} These properties guarantee that when $\bsplay\neq\zero$ it can be represented as 
\begin{equation} \label{eq:biaxial_splay_representation}
\bsplay=q(\n_1\otimes\n_1-\n_2\otimes\n_2),
\end{equation}
where $q$ is the \emph{positive} eigenvalue of $\bsplay$. This choice of sign for $q$ identifies (to within a sign) the eigenvectors $\n_1$ and $\n_2$ orthogonal to $\n$. We set $q=0$ when $\bsplay=\zero$. Since $\tr\bsplay^2=2q^2$, a useful identity follows from \eqref{eq:decomposition_gradient},
\begin{equation}
\label{eq:q_squared_expression}
2q^2=\tr(\gradn)^2+\frac12T^2-\frac12S^2.
\end{equation}

Selinger \cite{selinger:interpretation} has given compelling reasons to call $q$ the \emph{biaxial splay}; we shall adopt this name as well. He also convincingly argued that $T$ should be called the \emph{double} twist; however, here we shall stick to tradition and use the conventional name for it. Whenever $q>0$, the eigenvectors $\dframe$ of $\bsplay$ identify the \emph{distortion frame}. The four components of $\gradn$ in \eqref{eq:decomposition_gradient} are independent from one another; they identify four independent \emph{measures of distortion}, which we collect in $\measures$.

The first advantage of the novel decomposition of $\gradn$ in \eqref{eq:decomposition_gradient} is rewriting $\FF$ as the sum of four independent quadratics,
\begin{equation}\label{eq:Frank_energy_density_rewritten}
F_\mathrm{F}=\frac12(K_{11}-K_{24})S^2+\frac12(K_{22}-K_{24})T^2+\frac12K_{33}B^2+2K_{24}q^2,
\end{equation}
where $B^2:=\bend\cdot\bend$. This form of $\FF$ makes it immediate proving the conditions that render it positive definite,
\begin{equation}
\label{eq:ericksen_inequalities}
K_{11}-K_{24}>0,\quad K_{22}-K_{24}>0,\quad K_{33}>0,\quad K_{24}>0,
\end{equation}
also known as  Ericksen's inequalities \cite{ericksen:inequalities}.

The second advantage of \eqref{eq:decomposition_gradient} is to suggest an intriguing question \cite{selinger:interpretation}: Is it possible to fill space with a combination of uniform splay, twist, bend, and biaxial splay? In two space dimensions, the answer to this question depends on the Gaussian curvature of the surface on which the field $\n$ is defined. For a flat surface, apart from the trivial case of a constant $\n$, where both splay and bend are zero,\footnote{The twist is zero for all planar fields.} it is \emph{impossible} to construct a director field with non-zero uniform splay or  non-zero uniform bend \cite[p.\,320]{meyer:structural}. But this is possible for a surface of (constant) negative Gaussian curvature \cite{niv:geometric}.

Here we answer this question in three space dimensions. In Sec.~\ref{sec:uniform}, we introduce a definition of uniform distortions and prove that they are exhausted by only two families of fields. The explicit construction of such fields in Sec.~\ref{sec:heliconical} shows that they are Meyer's heliconical distortions \cite{meyer:structural}, which have recently been identified experimentally in the ground state of \emph{twist-bend} nematic phases \cite{cestari:phase}. In Sec.~\ref{sec:generalized}, we take advantage of the ground-state role played by uniform distortions in these phases to propose a simple elastic free-energy density that extends $\FF$ and has the potential to explain how the still mysterious twist-bend nematics can germinate out of ordinary nematics. Section~\ref{sec:conclusions} contains the conclusions of this work. Three closing appendices host some extra mathematical details.

\section{Uniform Distortions}\label{sec:uniform}
We have introduced in Sec.~\ref{sec:intro} the distortion frame $\dframe$, which can be defined for any sufficiently regular director field $\n$. Actually, the distortion frame is itself a field of frames, as in general it changes from place to place, and so do the components $(b_1,b_2)$ of the \emph{bend vector} $\bend$ expressed in the form
\begin{equation}\label{eq:b_components}
\bend=b_1\n_1+b_2\n_2,
\end{equation}
as well as $S$, $T$, and $q$. Seen from the distortion frame, the director field is locally characterized by the scalars $(S,T,b_1,b_2,q)$, which we call the \emph{distortion characteristics}.

Suppose that there is a director field $\n$ such that its distortion characteristics are the same everywhere, although the distortion frame may not be. For such a field, we could not tell where we are in space by sampling the local nematic distortion: we could not distinguish points with higher distortion (such as defects) from points with lower distortion. It is thus natural to call \emph{uniform} any such distortion. 

Clearly, the class of uniform distortions is \emph{not} empty: we know that any constant field $\n\equiv\n_0$ would obviously be uniform (but with no distortion). The very question is whether constant fields are indeed the \emph{only} uniform distortions. This question is answered for the \emph{negative} in this section, where we characterize all possible uniform distortions. This issue is intimately related to the possible nature of ground states for liquid crystals, and issue deferred to Sec.~\ref{sec:generalized} below. Here we assume $q>0$. The case $q=0$, for which the distortion frame $\dframe$ is undefined, will be treated in Sec.~\ref{sec:case}.

\subsection{Connectors}\label{sec:connectors}
The unit vectors in the distortion frame $\dframe$  it must satisfy the identities
\begin{subequations}\label{eq:distortion_frame_identities}
	\begin{gather}
	(\gradn_1)\trans\n_2+(\gradn_2)\trans\n_1=\zero,\label{eq:distortion_frame_identities_1}\\
	(\gradn_2)\trans\n+(\gradn)\trans\n_2=\zero,\label{eq:distortion_frame_identities_2}\\
	(\gradn)\trans\n_1+(\gradn_1)\trans\n=\zero,\label{eq:distortion_frame_identities_3}
	\end{gather}
	which stem from the mutual orthogonality of the vectors in a frame, and the identities
	\begin{equation}\label{eq:distortion_frame_ideintities_4}
	(\gradn_1)\trans\n_1=(\gradn_2)\trans\n_2=(\gradn)\trans\n=\zero,
	\end{equation}
\end{subequations}
which stem from having scaled to unity the length of the  eigenvectors of $\bsplay$. Identities \eqref{eq:distortion_frame_identities} combined together amount to represent the gradient of the vectors in the distortion frame as follows,
\begin{subequations}\label{eq:connectors}
	\begin{align}
	\gradn&=\n_1\otimes\con_1+\n_2\otimes\con_2,\label{eq:connectors_1}\\
	\gradn_1&=-\n\otimes\con_1+\n_2\otimes\cond,\label{eq:connectors_2}\\
	\gradn_2&=-\n\otimes\con_2-\n_1\otimes\cond,\label{eq:connectors_3}
	\end{align}
\end{subequations}
where $\con_1$, $\con_2$, and $\cond$ are vectors, which we call the \emph{connectors}. Both $\con_1$ and $\con_2$ are readily identified by the basic decomposition formula for $\gradn$ in \eqref{eq:decomposition_gradient}, which we reproduce here combined with \eqref{eq:biaxial_splay_representation} for the ease of the reader, 
\begin{equation}\label{eq:decomposition_gradient_reproduced}
\gradn=-\bend\otimes\n+\frac12T\W(\n)+\frac12S\Proj(\n)+q(\n_1\otimes\n_1-\n_2\otimes\n_2).
\end{equation}
A direct comparison between \eqref{eq:connectors_1} and \eqref{eq:decomposition_gradient_reproduced} yields
\begin{subequations}\label{eq:connectors_formulas}
	\begin{gather}
	\con_1=\left(\frac12S+q\right)\n_1-\frac12T\n_2-b_1\n,\label{eq:connectors_formula_1}\\
	\con_2=\frac12T\n_1+\left(\frac12S-q\right)\n_2-b_2\n,\label{eq:connectors_formula_2}
	\end{gather}
\end{subequations} 
where, according to \eqref{eq:b_components}, $b_1$ and $b_2$ are the components of the bend vector $\bend$ along $\n_1$ and $\n_2$, respectively. The third connector $\cond$ remains undetermined and will be derived in the following section to ensure that \eqref{eq:decomposition_gradient_reproduced} can be extended uniformly to the whole space.

\subsection{Compatibility Conditions}\label{sec:compatibility}
According to the definition given above, a uniform director field has all scalar distortion characteristics $(S,T,b_1,b_2,q)$ constant in space. For that to be the case, there must exist a connector $\cond$ such that both second gradients $\hess\n$ and $\hess\n_1$ be symmetric in the last two components, to ensure integrability in the whole space for both fields $\n$ and $\n_1$. Requiring the same condition for $\n_2$ would not be necessary, as once $\n$ and $\n_1$ are determined by integration of \eqref{eq:connectors_1} and \eqref{eq:connectors_2}, $\n_2$ is uniquely determined by setting $\n_2=\n\times\n_1$ and \eqref{eq:connectors_3} is entailed as a consequence.  

It follows from \eqref{eq:connectors_1} that 
\begin{equation}\label{eq:second_gradient_n}
\hess\n=\n\otimes\nabla\con_1-\n\otimes\con_1\otimes\con_1+\n_2\otimes\con_1\otimes\cond+\n_2\otimes\nabla\con_2-\n\otimes\con_2\otimes\con_2-\n_1\otimes\con_2\otimes\cond.
\end{equation}
This is a third-rank tensor, which is symmetric in the last two entries whenever the three second-rank tensors, $\n_1\cdot\hess\n$, $\n_e\cdot\hess\n$, and $\n\cdot\hess\n$, obtained saturating the first entry of $\hess\n$ with $\n_1$, $\n_2$, and $\n$, respectively, are all symmetric. Now, \eqref{eq:second_gradient_n} readily implies that 
\begin{subequations}\label{eq:second_gradient_components}
	\begin{eqnarray}
	\n_1\cdot\hess\n&=&\nabla\con_1-\con_2\otimes\cond,\label{eq:second_gradient_components_1}\\
	\n_2\cdot\hess\n&=&\con_1\otimes\cond+\nabla\con_2,\label{eq:second_gradient_components_2}\\
	\n\cdot\hess\n&=&-\con_1\otimes\con_1-\con_2\otimes\con_2.\label{eq:second_gradient_components_3}
	\end{eqnarray}
\end{subequations}
The last of these tensors is automatically symmetric, and so the integrability requirement for $\n$ amounts to the symmetry of the first two tensors. Keeping all distortion characteristics constant in \eqref{eq:connectors_formulas}, we see that
\begin{subequations}\label{eq:connector_gradients}
	\begin{eqnarray}
	\nabla\con_1&=&\left(\frac12S+q\right)\nabla\n_1-\frac12T\nabla\n_2-b_1\nabla\n,\label{eq:connector_gradient_1}\\
	\nabla\con_2&=&\frac12T\nabla\n_1+\left(\frac12S-q\right)\nabla\n_2-b_2\nabla\n,\label{eq:connector_gradient_2}
	\end{eqnarray}
\end{subequations}
where $\gradn$, $\gradn_1$, and $\gradn_2$, via \eqref{eq:connectors} and \eqref{eq:connectors_formulas}, are meant to be expressed again in terms of $(S,T,b_1,b_2,q)$ and the still unknown components $(d_1,d_2,d_3)$ of $\cond$ in the frame $\dframe$. Like the components  of the connectors $\con_1$ and $\con_2$, $d_1,d_2,d_3$ are also taken to be uniform in space.\footnote{Clearly, like the other two connectors, $\cond$ fails in general to be uniform in space.}

Requiring the first two tensors in \eqref{eq:second_gradient_components} to be symmetric leads us to six scalar equations in the eight unknowns $(S,T,b_1,b_2,q, d_1,d_2,d_3)$. After some rearrangements, they read as follows,
\begin{subequations}\label{eq:n_symmetry_equations}
	\begin{eqnarray}
	2qd_1&=&b_1T,\label{eq:n_symmetry_equations_1}\\
	b_2d_1&=&b_1^2-\tfrac14\left(T^2-S^2\right)+q(S+q),\label{eq:n_symmetry_equations_2}\\
	2qd_3-b_2d_2&=&\tfrac12ST-b_1b_2,\label{eq:n_symmetry_equations_3}\\
	2qd_2&=&-b_2T,\label{eq:n_symmetry_equations_4}\\
	2qd_3+b_1d_1&=&-b_1b_2-\tfrac12ST,\label{eq:n_symmetry_equations_5}\\
	b_1d_2&=&-b_2^2+\tfrac14\left(T^2-S^2\right)+q(S-q),\label{eq:n_symmetry_equations_6}
	\end{eqnarray}
\end{subequations}
where we have isolated the terms linear in the $d$'s. Since $q>0$, it readily follows from \eqref{eq:n_symmetry_equations_1} and \eqref{eq:n_symmetry_equations_4} that
\begin{subequations}\label{eq:d_solutions}
	\begin{equation}\label{eq:d_1_d_2_solutions}
	d_1=\frac{b_1T}{2q}\quad\text{and}\quad d_2=-\frac{b_2T}{2q}.
	\end{equation}
	Inserting these in the remaining equations \eqref{eq:n_symmetry_equations}, we obtain the following expression for $d_3$,
	\begin{equation}\label{eq:d_3_solution}
	d_3=-\frac{1}{4q}\left(2b_1b_2+\frac{T}{2q}(b_1^2+b_2^2)\right),
	\end{equation}
\end{subequations}
supplemented by the equations
\begin{subequations}\label{eq:supplentary_equations}
	\begin{equation}\label{eq:supplementary_equation_1}
	\frac{b_1b_2T}{2q}=\frac12\left(b_1^2+b_2^2-\frac12(T^2-S^2)+2q^2\right)
	\end{equation}
	and
	\begin{equation}\label{eq:supplementary_equation_2}
	S=\frac{1}{2q}(b_1^2-b_2^2).
	\end{equation}
\end{subequations}
By combining together equations \eqref{eq:supplentary_equations}, we finally solve for $T$, arriving at the following two roots,
\begin{subequations}\label{eq:T_roots}
	\begin{eqnarray}
	T_1&=&\frac{1}{2q}(b_1-b_2)^2+2q,\label{eq:T_1}\\
	T_2&=&-\frac{1}{2q}(b_1+b_2)^2-2q.\label{eq:T_2}
	\end{eqnarray}	
\end{subequations}
Making use of these latter in \eqref{eq:d_solutions} and \eqref{eq:supplementary_equation_2}, we conclude that the symmetry requirement for the tensors in \eqref{eq:second_gradient_components_1} and \eqref{eq:second_gradient_components_2} are satisfied by letting $S$ and $T$ be related to $(b_1,b_2,q)$ through \eqref{eq:supplementary_equation_2} and \eqref{eq:T_roots}. Thus, there are two  families of distortion characteristics compatible with the symmetry of $\hess\n$: they differ by the sign of the twist $T$, being $T_1>0$ and $T_2<0$ (since $q>0$), and are parameterized by $(b_1,b_2,q)$, which remain free; the components of the connector $\cond$ are correspondingly delivered by \eqref{eq:d_solutions}.

Starting from \eqref{eq:decomposition_gradient_reproduced}, we have ensured that $\hess\n$ is symmetric, but this is not enough to guarantee that the complete frame $\dframe$ can be extended through the whole space keeping \eqref{eq:decomposition_gradient_reproduced} valid. To do this, starting from \eqref{eq:connectors_2}, we also need to ensure that $\hess\n_1$ stay symmetric when the connectors obey \eqref{eq:connectors_formulas} and \eqref{eq:d_solutions}. 

Retracing our steps above, with the aid of  \eqref{eq:connectors}, we now write
\begin{equation}\label{eq:second_gradient_n_1}
\hess\n_1=-\n\otimes\nabla\con_1-\n_2\otimes\con_1\otimes\con_2-\n_1\otimes\con_1\otimes\con_1-\n_2\otimes\nabla\cond-\n_1\otimes\cond\otimes\cond-\n\otimes\cond\otimes\con_2
\end{equation}
and find the analogs of \eqref{eq:second_gradient_components},
\begin{subequations}\label{eq:second_gradient_components_n_1}
	\begin{eqnarray}
	\n_1\cdot\hess\n_1&=&-\con_1\otimes\con_1-\cond\otimes\cond,\label{eq:second_gradient_components_n_1_1}\\
	\n_2\cdot\hess\n_1&=&-\con_1\otimes\con_2+\nabla\cond,\label{eq:second_gradient_components_n_1_2}\\
	\n\cdot\hess\n_1&=&-\nabla\con_1-\cond\otimes\con_2,\label{eq:second_gradient_components_n_1_3}
	\end{eqnarray}	
\end{subequations}
where $\con_1$ and $\con_2$ are as in \eqref{eq:connectors_formulas} and the components of $\cond$ in the frame $\dframe$ are to be given by \eqref{eq:d_solutions}. Clearly, the  tensor in  \eqref{eq:second_gradient_components_n_1_1} is already symmetric. The symmetry condition for the tensors in \eqref{eq:second_gradient_components_n_1_2} and \eqref{eq:second_gradient_components_n_1_3} amounts to the following set of scalar equations,
\begin{subequations}\label{eq:n_1_symmetry_equations}
	\begin{eqnarray}
	d_1^2+d_2^2+Td_3&=&-\tfrac14T^2-\tfrac14S^2+q^2,\label{eq:n_1_symmetry_equations_1}\\d_3(d_2+b_1)-(\tfrac{1}{2}S+q)d_1-\tfrac12Td_2&=&b_2(\tfrac12S+q)-\tfrac12b_1T,\label{eq:n_1_symmetry_equations_2}\\
	d_3(d_1-b_2)-\tfrac12Td_1+(\tfrac12S-q)d_2&=&b_1(\tfrac12S-q)+\tfrac12b_2T,\label{eq:n_1_symmetry_equations_3}\\
	2qd_1&=&b_1T,\label{eq:n_1_symmetry_equations_4}\\
	b_2d_1&=&b_1^2+(\tfrac12S+q)^2-\tfrac14T^2,\label{eq:n_1_symmetry_equations_5}\\
	2qd_3&=&-\tfrac{1}{2q}b_2^2T-b_1b_2+\tfrac12ST,\label{eq:n_1_symmetry_equations_6}
	\end{eqnarray}
\end{subequations}
where again the terms in the $d$'s (though no longer all linear) have been isolated from the others.

We see that \eqref{eq:n_1_symmetry_equations_4} is nothing but \eqref{eq:n_symmetry_equations_1}, and \eqref{eq:n_1_symmetry_equations_6} reduces to \eqref{eq:d_3_solution}, as soon as we make use of \eqref{eq:supplementary_equation_2}. Similarly, use of \eqref{eq:d_1_d_2_solutions} and \eqref{eq:supplementary_equation_2} in \eqref{eq:n_1_symmetry_equations_5} turns the latter into an identity. As for the remaining equations \eqref{eq:n_1_symmetry_equations}, \eqref{eq:d_solutions} transforms \eqref{eq:n_1_symmetry_equations_1} into
\begin{equation}\label{eq:intermediate}
\left(\frac{T^2}{(2q)^2}-1\right)\left(2q^2+\frac12(b_1^2+b_2^2)\right)=0,
\end{equation}
which implies that
\begin{equation}\label{eq:T_equation}
T^2=(2q)^2
\end{equation}
This, combined with the tow variants in \eqref{eq:T_roots}, leaves us with the alternative
\begin{equation}\label{eq:B-alternative}
b_1=b_2\quad\text{or}\quad b_1=-b_2.
\end{equation}
In both instances, \eqref{eq:supplementary_equation_2} implies that $S=0$, and direct inspection of \eqref{eq:n_1_symmetry_equations_2} and \eqref{eq:n_1_symmetry_equations_3} shows that they are then identically satisfied. 

Recapitulating, we conclude that there exist only \emph{two} families of uniform director fields, according to the definition given in this work. They are classified as follows:
\begin{subequations}\label{eq:families}
	\begin{eqnarray}
	S&=&0,\quad T=2q,\quad b_1=b_2=b,\label{eq:families_1}\\
	S&=&0,\quad T=-2q,\quad b_1=-b_2=b,\label{eq:families_2}
	\end{eqnarray}
\end{subequations}
where $q>0$ and $b$ are arbitrary scalar parameters. Correspondingly, the connectors $(\con_1,\con_2,\cond)$ are given by 
\begin{subequations}\label{eq:families_connectors}
	\begin{eqnarray}
	\con_1&=&\con_2=q\n_1-q\n_2-b\n,\quad \cond=b\n_1-b\n_2-\tfrac{b^2}{q}\n,\label{eq:families_connectors_1}\\
	\con_1&=&-\con_2=q\n_1+q\n_2-b\n,\quad \cond=-b\n_1-b\n_2+\tfrac{b^2}{q}\n.\label{eq:families_connectors_2}
	\end{eqnarray}
\end{subequations}
The connection between equations \eqref{eq:families} and \eqref{eq:families_connectors} and the \emph{heliconical} director distortions is illustrated in the following section. Our development above has shown that they are the only possible families of uniform distortions, each distinguished by the sign of the twist.

\subsection{Case $q=0$}\label{sec:case}
In the above analysis $q$ was \emph{positive}. When $q=0$, the distortion frame $\dframe$ is no longer defined, because $\bsplay=\zero$, but the notion of uniform distortion still makes sense. Here we show how to extend its definition to this case.

First, let also $B=0$. Then \eqref{eq:decomposition_gradient_reproduced} reduces to 
\begin{equation}\label{eq:decomposition_gradient_q_and_B_zero}
\gradn=\frac12T\W(\n)+\frac12S\Proj(\n).
\end{equation} 
A distortion is uniform only if there is a solution of \eqref{eq:decomposition_gradient_q_and_B_zero} with both $S$ and $T$ constant in space. It readily follows from \eqref{eq:decomposition_gradient_q_and_B_zero} that
\begin{subequations}\label{eq:q_zero_compatibility}
\begin{eqnarray}
\n\cdot\nabla^2\n&=&-\frac14(S^2+T^2)\Proj(\n),\label{eq:q_zero_compatibility_n}\\
\e\cdot\nabla^2\n&=&\frac14(T^2-S^2)\n\otimes\e+\frac14ST\W(\e)+\frac14ST(\e_\perp\otimes\n+\n\otimes\e_\perp),\label{eq:q_zero_compatibility_e}
\end{eqnarray}
\end{subequations}
where $\e$ is any unit vector orthogonal to $\n$, $\e_\perp:=\n\times\e$, and $\W(\e)$ is the skew-symmetric tensor associated with $\e$. While the tensor in \eqref{eq:q_zero_compatibility_n} is always symmetric, the tensor in \eqref{eq:q_zero_compatibility_e} is so only if $T^2-S^2=0$ and $ST=0$, which imply $\gradn\equiv\zero$, that is, $\n$ is itself trivially uniform. 

If $B\neq0$, we can formally define a distortion frame $\dframe$ by letting $\bend=B\n_1$ and $\n_2=\n\times\n_1$. Then the analysis in Secs.~\ref{sec:connectors} and \ref{sec:compatibility} go through unchanged, provided we set $b_1=B$, $b_2=0$, $q=0$ in \eqref{eq:n_symmetry_equations}. It is a simple matter to check that equations \eqref{eq:n_symmetry_equations} would then turn incompatible in $(S,T,B)$, for arbitrary $(d_1,d_2,d_3)$.

The conclusion is that for $q=0$ the only uniform distortion is the trivial uniform field.
	
\section{Heliconical Distortions}\label{sec:heliconical}
In this section, we show how to integrate \eqref{eq:decomposition_gradient_reproduced} when the distortion characteristics are specified as in either of equations \eqref{eq:families}. This will allow us to establish that the most general uniform distortion is a \emph{heliconical} director field. We shall also show how the free parameters $(q,b)$ in \eqref{eq:families} are related to the pitch $P$ and the conical angle $\alpha$ that identify a heliconical director field.

First, we consider a trajectory $\traj$ in space parameterized in its arc-length $s$, and imagine to follow the distortion frame $\dframe$ as its origin progresses along $\traj$. In complete analogy with rigid body dynamics, if we interpret $s$ as time, we can say that there must be a vector $\spin$ such that 
\begin{equation}\label{eq:distortion_frame_evolution}
\n_1'=\spin\times\n_1,\quad\n_2'=\spin\times\n_2,\quad\n'=\spin\times\n,
\end{equation} 
where a prime $'$ denotes differentiation along the path $\traj$ (that is, with respect to $s$). Letting $\e$ denote the unit tangent vector to $\traj$, we have that 
\begin{equation}\label{eq:distortion_frame_evolution_rewritten}
\n_1'=(\gradn_1)\e,\quad\n_2'=(\gradn_2)\e,\quad\n'=(\gradn)\e,
\end{equation}
and comparing \eqref{eq:distortion_frame_evolution} and \eqref{eq:distortion_frame_evolution_rewritten} with \eqref{eq:connectors}, we easily see that $\spin$ depends linearly on $\e$, $\spin=\Spin\e$, where $\Spin$ is a tensor that can be expressed in terms of the connectors as
\begin{equation}\label{eq:Omega_formula}
\Spin=\n_2\otimes\con_1-\n_1\otimes\con_2+\n\otimes\cond.
\end{equation}

Second, we ask a question. Is there any eigenvector $\e$ of $\Spin$? The answer to this question is relevant to the geometric interpretation of uniform distortions. Were $\e$ an eigenvector  of $\Spin$, $\e'=\spin\times\e=\zero$; as a consequence, $\e$ would be constant along $\traj$ and the latter would be a straight line. Thus, the eigenvectors of $\Spin$, if they exist, identify directions in space around which the distortion frame $\dframe$ precesses with a winding rate (pitch) prescribed by the corresponding eigenvalue $\lambda$.

It follows from \eqref{eq:Omega_formula} that an eigenpair $(\lambda,\e)$ of $\Spin$ must satisfy the equation
\begin{equation}\label{eq:Omega_eigenpair}
(\con_1\cdot\e)\n_2-(\con_2\cdot\e)\n_1+(\cond\cdot\e)\n=\lambda\e.
\end{equation}
For $\con_1$, $\con_2$, and $\cond$ given by \eqref{eq:families_connectors_1}, corresponding to the first family of uniform distortions obtained in Sec.~\ref{sec:compatibility}, equation \eqref{eq:Omega_eigenpair} reduces to the following three scalar linear equations,
\begin{subequations}\label{eq:Omega_eigenpair_scalar}
\begin{eqnarray}
qe_1-qe_2-be_3&=&\lambda e_2,\label{eq:Omega_eigenpair_scalar_1}\\
qe_1-qe_2-be_3&=&-\lambda e_1,\label{eq:Omega_eigenpair_scalar_2}\\
be_1-be_2-\tfrac{b^2}{q}e_3&=&\lambda e_3,\label{eq:Omega_eigenpair_scalar_3}
\end{eqnarray}
\end{subequations}
for the components $(e_1,e_2,e_3)$ of $\e$ in the frame $\dframe$. Requiring the system \eqref{eq:Omega_eigenpair_scalar} to have zero determinant (which is the solvability condition for $\e$), we obtain the \emph{secular equation} for $\lambda$,
\begin{equation}\label{eq:lambda_secular_equation}
\lambda^2\left(\lambda+2q+\frac{b^2}{q}\right)=0,
\end{equation}
which has three real roots,
\begin{equation}\label{eq:roots_secular_equation}
\lambda_1=\lambda_2=0\quad\text{and}\quad\lambda_3=-2q-\frac{b^2}{q}.
\end{equation}
The (unoriented) eigenvector $\e$ corresponding to $\lambda_3$ has components
\begin{equation}\label{eq:eigenvector_3_components}
e_1=\mp\frac{q}{\sqrt{b^2+2q^2}},\quad e_2-e_1,\quad e_3=\pm\frac{b}{\sqrt{b^2+2q^2}},
\end{equation}
whereas the components $(\widehat{e}_1,\widehat{e}_2,\widehat{e}_3)$ of the eigenvectors $\widehat{\e}$ corresponding to the eigenvalues $\lambda_{1,2}$ are the solutions to the equation
\begin{equation}\label{eq:eigenvectors_zero_eigenvalue}
q\widehat{e}_1-q\widehat{e}_2-b\widehat{e}_3=0.
\end{equation}
Contrasting \eqref{eq:eigenvectors_zero_eigenvalue} with \eqref{eq:eigenvector_3_components}, we immediately see that $\widehat{\e}$ is any unit vector orthogonal to $\e$.

Geometrically, this means that the distortion frame $\dframe$ precesses anti-clockwise (because $\lambda_3<0$) along $\e$ (whatever orientation we take for the latter), turning completely round over the length of a \emph{pitch},
\begin{equation}\label{eq:pitch}
P=\frac{2\pi}{|\lambda_3|}=\frac{2\pi q}{b^2+2q^2},
\end{equation}
whereas it remains unchanged in all directions orthogonal to $\e$. The nematic field thus described is nothing but the \emph{heliconical} distortion first hypothesized by Meyer~\cite[p.\,320]{meyer:structural} and recently recognized as being the fingerprint of the \emph{twist-bend} liquid crystal phase, the newest nematic phase, discovered only in 2011 \cite{cestari:phase}.\footnote{We shall say more about twist-bend nematics in Sec.~\ref{sec:achiral} below.} The nematic director $\n$ makes a fixed \emph{cone} angle $\alpha$ with the rotation axis $\e$, which is also called the \emph{helix} axis. A glance at \eqref{eq:eigenvector_3_components} suffices to see that the least determination of $\alpha$ satisfies the equation\footnote{Incidentally, both formulas \eqref{eq:pitch} and \eqref{eq:cone_angle_formula} agree with the explicit, geometric representation of a heliconical field, such as that embodied by equations (2) through (4) of \cite{virga:double-well}.}
\begin{equation}\label{eq:cone_angle_formula}
\cos\alpha=\frac{|b|}{\sqrt{b^2+2q^2}}.
\end{equation}
In Appendix~\ref{sec:construction} we show in details how to construct the heliconical nematic fields $\n$ corresponding to the eigenvalues and eigenvectors of $\Spin$. There, it will also become apparent why the orientation of the eigenvector $\e$ is immaterial to this construction.

Although $\e$ can be chosen with either of the signs in \eqref{eq:eigenvector_3_components}, it may be useful to select conventionally an orientation that would guide the eye ad avoid unnecessary confusion. Our choice is to orient the helix axis $\e$ in such a way that the director $\n$ makes an acute angle with it. By \eqref{eq:eigenvector_3_components}, we see that this orientation depends uniquely on the sign of $b$,\footnote{Of course, this choice relies on having chosen an orientation also for $\n$, which for uniform fields turns out to be always possible.}
\begin{equation}\label{eq:eigenvector_3_oriented_components}
e_1=-\sgn(b)\frac{q}{\sqrt{b^2+2q^2}},\quad e_2=-e_1,\quad e_3=\frac{|b|}{\sqrt{b^2+2q^2}}.
\end{equation}

The family of uniform distortions in \eqref{eq:families_2} can be treated in precisely the same way. The only difference with respect to the one in \eqref{eq:families_1} is that $\lambda_3$ is now positive,
\begin{equation}\label{eq:lambda_3_new}
\lambda_3=2q+\frac{b^2}{q},
\end{equation}
so that the distortion frame $\dframe$ precesses clockwise along the helix axis $\e'$, which differs from $\e$: its components are
\begin{equation}\label{eq:eigenvector_e_3_prime_components}
e_1'=\pm\frac{q}{\sqrt{b^2+2q^2}},\quad e_2'=e_1',\quad e_3'=\mp\frac{b}{\sqrt{b^2+2q^2}}.
\end{equation}
Adopting for the orientation of $\e'$ the same convention introduced for $\e$, we replace \eqref{eq:eigenvector_e_3_prime_components} with
\begin{equation}\label{eq:eigenvector_e_3_prime_oriented_components}
e_1'=\sgn(b)\frac{q}{\sqrt{b^2+2q^2}},\quad e_2'=e_1',\quad e_3'=\frac{|b|}{\sqrt{b^2+2q^2}}.
\end{equation}
Comparing \eqref{eq:eigenvector_e_3_prime_oriented_components} with \eqref{eq:eigenvector_3_oriented_components}, we see that the oriented helix axes of the two families of uniform distortions (with opposite twists) are such that
\begin{equation}\label{eq:helix_axes}
e_1e_1'+e_2e_2'=0,\quad e_3e_3'=\frac{b^2}{b^2+2q^2}\geqq0.
\end{equation}
This shows that the projections of $\e$ and $\e'$ on the plane orthogonal to $\n$ are perpendicular to one another. Moreover, upon reversing the sign of $b$ both these projections get reversed. Finally, both pitch $P$ and cone angle $\alpha$ are delivered by the same formulas \eqref{eq:pitch} and \eqref{eq:cone_angle_formula}, respectively.

\begin{figure}[h]
\centering
\begin{subfigure}[b]{0.3\textwidth}
	\centering
	\includegraphics[clip,width=\textwidth]{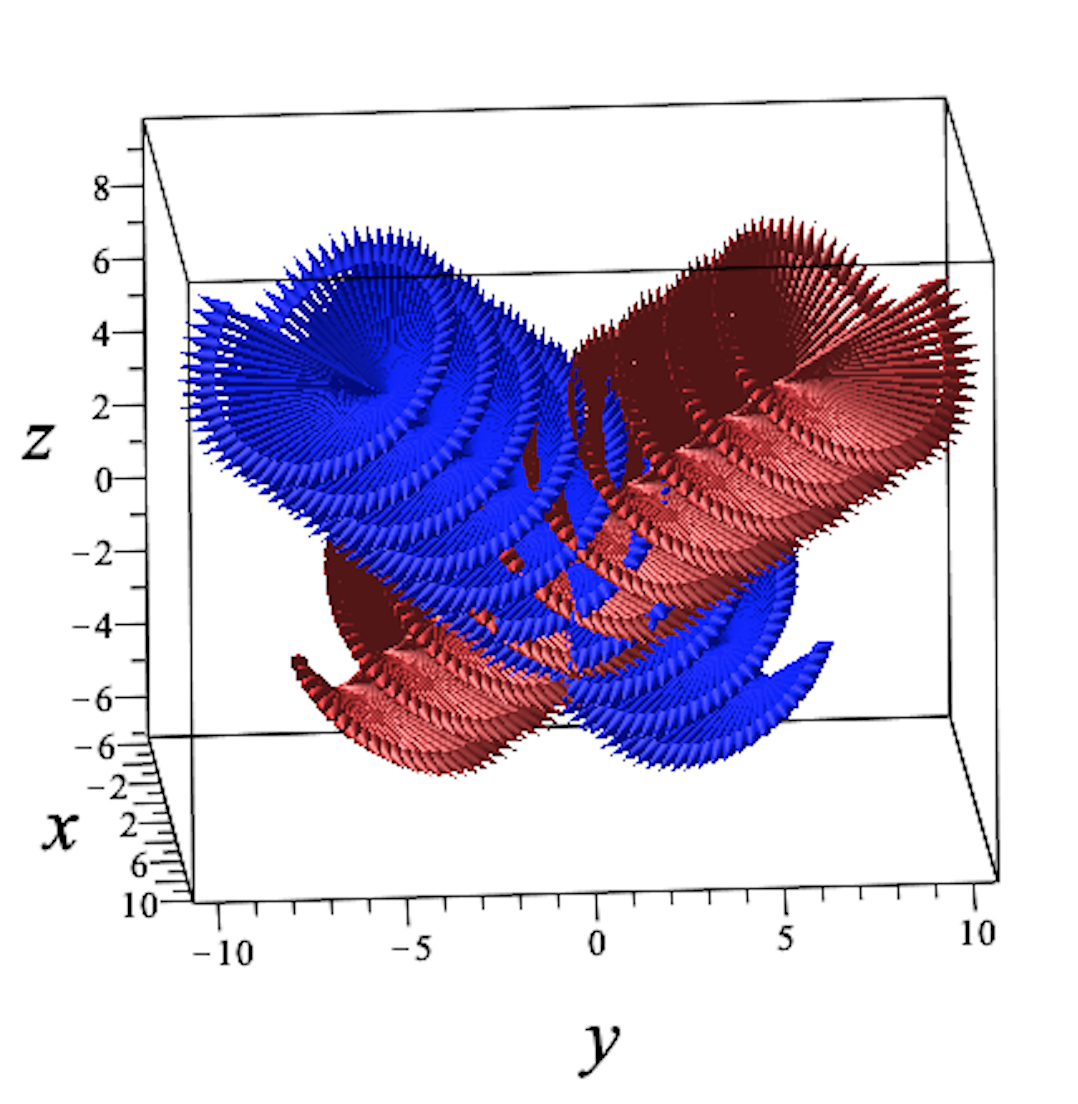}
	\caption{$\qquad b/q=-1$}
\end{subfigure}
\begin{subfigure}[b]{0.3\textwidth}
	\centering
	\includegraphics[clip,width=\textwidth]{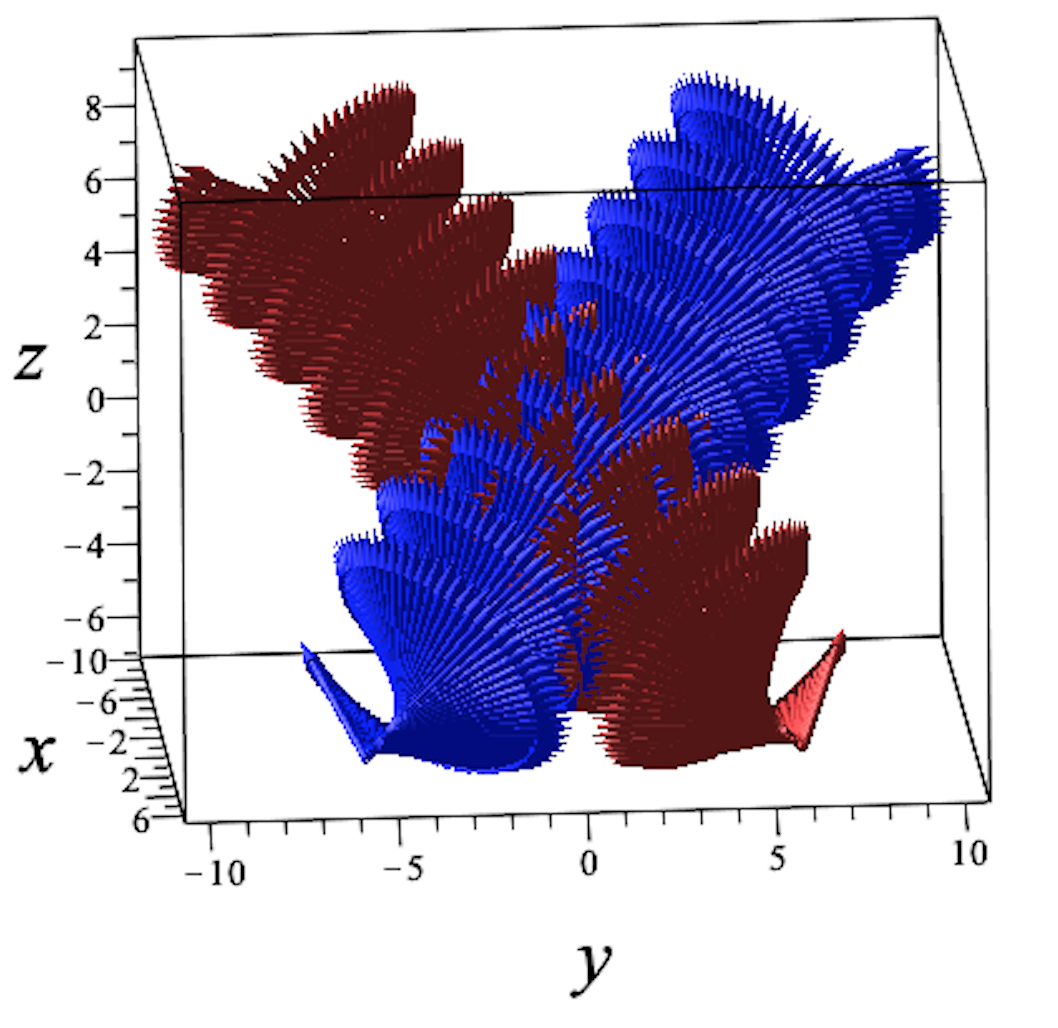}
	\caption{$\qquad b/q=1$}
\end{subfigure}
\caption{(Color online) The heliconical nematic fields with negative (blue) and positive (brown) eigenvalue $\lambda_3$, as delivered by \eqref{eq:roots_secular_equation} and \eqref{eq:lambda_3_new},  respectively. Panels (a) and (b) also illustrate the symmetries of the helix axes embodied by \eqref{eq:eigenvector_3_oriented_components} and \eqref{eq:eigenvector_e_3_prime_oriented_components}. The blue field (for which $\lambda_3=-3q$) precesses counter-clockwise around the helix axis, whereas the brown field (for which $\lambda_3=3q$) precesses clockwise around the helix axis.}
\label{fig:heliconicals}
\end{figure}

Figure~\ref{fig:heliconicals} illustrates a three-dimensional representation of the heliconical fields in the two families \eqref{eq:families}. 
In Fig.~\ref{fig:heliconicals}, the frame $(\e_x,\e_y,\e_z)$ is chosen so as to coincide with the distortion frame $\dframe$ at the origin (where also $s=0$).\footnote{See Appendix~\ref{sec:construction} for more details.} Both $b$ and $q$ have the same physical dimensions (the inverse of a length). For representative purposes, here we rescale $b$ to $q$.

\begin{figure}[h]
	\centering
	\includegraphics[clip,width=.3\textwidth]{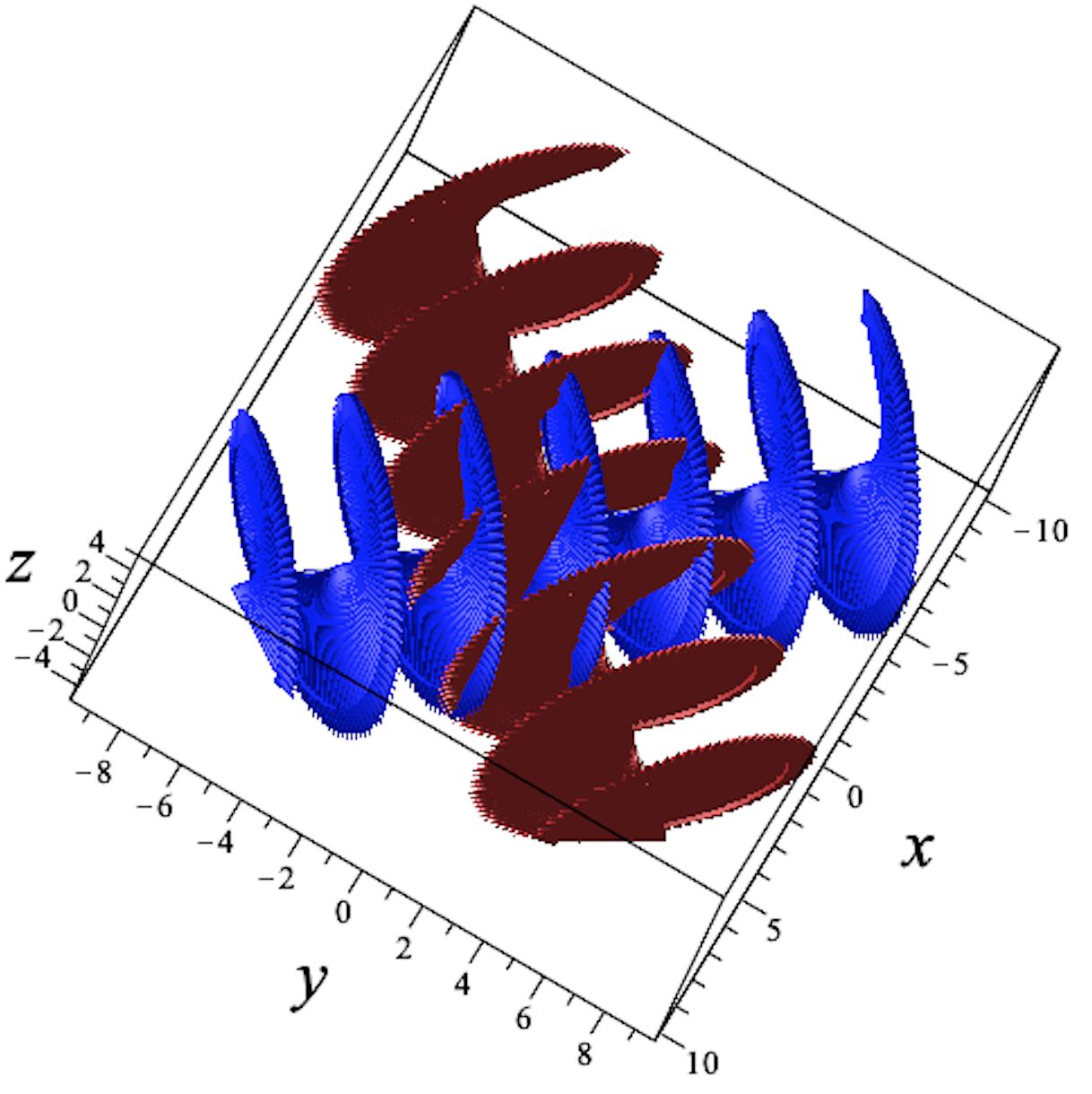}
	\caption{(Color online) The same heliconical director fields as in Fig.~\ref{fig:heliconicals}, but for $b=0$. The two helix axes are perpendicular to one another, as prescribed by \eqref{eq:b_0_axes}, and the cone angle takes on the limiting value $\alpha=\pi/2$.}
	\label{fig:heliconical_b_0}
\end{figure}

It is perhaps worth recalling that for $b=0$ the heliconical fields in Fig.~\ref{fig:heliconicals} reduce to the two variants of the single twist characteristic of the ground state of chiral nematics, for which
\begin{equation}\label{eq:chiral_nematics}
\alpha=\frac\pi2\quad\text{and}\quad P=\frac{\pi}{q}=\frac{2\pi}{|T|}.
\end{equation}
In \eqref{eq:chiral_nematics}, as in the general cases \eqref{eq:families}, it is not only the sign of $T$ that distinguishes the two chiral variants of the uniform distortions. They also have different helix axes. It follows from \eqref{eq:eigenvector_3_components} and \eqref{eq:eigenvector_e_3_prime_components} that their components in the frame $\dframe$ are given by
\begin{equation}\label{eq:b_0_axes}
\begin{split}
e_1=-\frac{1}{\sqrt{2}},\quad e_2=\frac{1}{\sqrt{2}},\quad e_3=0,\\
e_1'=\frac{1}{\sqrt{2}},\quad e_2'=\frac{1}{\sqrt{2}},\quad e_3'=0,\\
\end{split}
\end{equation}
so that, in accordance with \eqref{eq:helix_axes}, $\e\cdot\e'=0$. This limiting case is illustrated in Fig.~\ref{fig:heliconical_b_0}.

\section{Generalized Elasticity}\label{sec:generalized}
We have already seen how two families of heliconical distortions with opposite twists (including the limiting case of zero bend) represent the totality of uniform director distortions that can fill the whole space. Any other director field would be geometrically frustrated and become by necessity non-uniform, if requested to occupy the whole space. It is interesting to see whether one could easily construct an elastic theory that penalizes the departures from a selected uniform field in one of the families \eqref{eq:families}.

Thus we would generalize (in one of many possible ways) the classical elastic theory of Frank, by replacing the ground state where $\n$ is the same in the whole space with one or more members of the uniform families \eqref{eq:families}. Since only one of the distortion characteristics $(S,T,b_1,b_2,q)$ vanishes generically in the uniform families, namely $S$, a quadratic theory, such as Frank's, is no longer sufficient.

As lucidly recalled in \cite{selinger:interpretation}, there are essentially two avenues towards a higher-order theory, that is, to allow  either for higher spatial derivatives of $\n$ in the elastic free-energy density or for higher powers of its spatial gradient.\footnote{A hybrid approach has been proposed in \cite{lelidis:nonlinear} on the basis of a molecular derivation of the phenomenological free energy. There, the order of spatial derivatives and their powers are balanced according to a criterion motivated by a molecular model.} Here we shall follow the latter approach.

In this section, we shall only consider an achiral scenario, as it seems that phases with such a  ground state have already been identified experimentally. We shall rely on the construction of an appropriate double-well elastic free-energy density. Before doing so, we sketch the basic ingredients of the theory and the invariance properties that we require. 

As made clear by the decomposition of $\gradn$ in \eqref{eq:decomposition_gradient}, for a given $\n$, the measures of distortions are $\measures$, namely, a scalar, a pseudo-scalar, a vector, and a tensor, respectively. A further pseudo-vector and a further vector can be built starting from the measures of distortions; these are $\n\times\bend$ and $\bsplay\bend$, respectively. 

The \emph{nematic} symmetry requires that any physically significant scalar must be invariant under the transformation of $\n$ into $-\n$. Here is how the measures of distortion and their derived vector and pseudo-vector behave under this transformation:
\begin{equation}
\label{eq:nematic_symmetries}
S\to-S,\quad T\to T,\quad\bend\to\bend,\quad \bsplay\to-\bsplay,\quad\n\times\bend\to-\n\times\bend,\quad\bsplay\bend\to-\bsplay\bend.
\end{equation}
Similarly, the central inversion of space produces the following changes,
\begin{equation}
\label{eq:inversion_symmetries}
S\to S,\quad T\to -T,\quad\bend\to\bend,\quad\bsplay\to\bsplay,\quad\n\times\bend=-\n\times\bend,\quad\bsplay\bend\to\bsplay\bend.
\end{equation}
Thus, keeping in mind that $\tr\bsplay=\tr\bsplay^3=0$, we collect all generating monomials (to be multiplied up to the fourth power in $\gradn$) in the list
\begin{equation}
\label{eq:list_1}
\left\{S^2, T^2, B^2, \tr\bsplay^2, S\bend\cdot\bsplay\bend, T\bend\cdot\bsplay(\n\times\bend) \right\},
\end{equation}
which, being invariant under the combined action of \eqref{eq:nematic_symmetries} and \eqref{eq:inversion_symmetries}, applies to achiral nematics.

Lists such as \eqref{eq:list_1} are not completely new in the literature. The first three members of \eqref{eq:list_1} feature, for example, in the recent papers \cite{lelidis:nonlinear,barbero:fourth-order}, but the mixed quartic invariants involving three out the four measures of distortion appear to be new.

By use of \eqref{eq:biaxial_splay_representation} and \eqref{eq:b_components}, we easily see that 
\begin{subequations}\label{eq:identity_first_second}
	\begin{eqnarray}
	\bend\cdot\bsplay\bend&=&q(b_1^2-b_2^2),\label{eq:identity_first}\\
	\bend\cdot\bsplay(\n\times\bend)&=&-2qb_1b_2\label{eq:identity_second}.
	\end{eqnarray}
\end{subequations}
Similarly, we obtain
\begin{equation}\label{eq:identity_third}
(\n\times\bend)\cdot\bsplay(\n\times\bend)=q(b_2^2-b_1^2)=-\bend\cdot\bsplay\bend,
\end{equation} 
which shows how the invariant $(\n\times\bend)\cdot\bsplay(\n\times\bend)$ would be redundant in \eqref{eq:list_1}.

While $\tr\bsplay^2=2q^2$ can be directly expressed in terms of the invariants of $\gradn$ via \eqref{eq:q_squared_expression}, slightly more labor is required for the quartic invariants in \eqref{eq:list_1}. Use of \eqref{eq:identity_first_second}, \eqref{eq:identity_1}, and \eqref{eq:identity_2} (see Appendix~\ref{sec:identities}) leads us to the following expressions
\begin{subequations}\label{eq:identity_invariants}
\begin{eqnarray}
S\bend\cdot\bsplay\bend&=&T\bend\cdot(\gradn)\bend-\frac12S^2T^2,\label{eq:identity_invariants_first}\\
T\bend\cdot\bsplay(\n\times\bend)&=&T\curl\n\cdot(\gradn)\bend+\frac12T^2B^2.\label{eq:identity_invariants_second}
\end{eqnarray}	
\end{subequations}

In the remaining of this section, we shall consider an elastic free-energy density built from the members of \eqref{eq:list_1}.

\subsection{Generalized Achiral Nematics}\label{sec:achiral}
\emph{Twist-bend} nematics (\NTB) have been intensely studied in the past decade. This paper is not focused on these new phases, but we can hardly escape from them, as their ground state happens to be the  uniform distortion that a director field can generically have.

A great deal of theories and models have been put forward to explain how a twist-bend phase germinates out of ordinary nematics. Allegedly, the first elastic theory was proposed by Dozov~\cite{dozov:spontaneous}, who used higher derivatives in the free energy to counterbalance the instability produced in Frank's energy by a negative bend constant $K_{33}$. Other elastic theories, with different features and perspectives can be found in \cite{virga:double-well,barbero:elastic,lelidis:nonlinear,barbero:fourth-order}. Phenomenological Landau theories \cite{shamid:statistical},\cite{kats:landau,longa:modulated,aliev:helicoidal} and molecular field theories \cite{greco:molecular,tomczyk:twist-bend,osipov:effect,vanakaras:molecular} are also available, as well as accurate reviews \cite{panov:twist-bend,mandle:dependency}.

The twist-bend ground state is two-fold; it consists of two members (with opposite twist) taken from the heliconical families \eqref{eq:families}. Since the nematogenic molecules that comprise a \NTB\ phase are \emph{not} chiral, the two variants with opposite macroscopic chirality  are equally present in the phase and must be accounted for by an elastic theory. This is indeed the only example of spontaneous chiral symmetry breaking known in a fluid in the absence of spatial order \cite{copic:nematic}.\footnote{The modulated arrangement in a \NTB\ phase is  \emph{not} accompanied by a mass density wave \cite{chen:chiral}.}

Many experimental studies have claimed the existence of the \NTB\ phase in a number nematogenic systems with various molecular motifs \cite{chen:chiral,chen:twist-bend,borshch:nematic,gorecka:short,paterson:understanding,salamonczyk:structure,tuchband:distinct,zhu:resonant}. These studies agree in showing that the pitch of the modulated nematic structure, which indeed exhibits both chiralities, fall in the nanometric range. Strictly speaking, this would make it questionable to use a phenomenological elastic theory to explain the \NTB\ phase.
We shall, however, entertain the theoretical possibility that an elastic free-energy density quadratic in $\gradn$ could be minimized by both chiral variants of the uniform families \eqref{eq:families}.

We shall not consider the most general elastic free energy with the desired property; we shall be contented with a minimalistic approach that produces the simplest instance of such an energy. Since the putative minimizers in \eqref{eq:families} are characterized by having $b_1=b_2$ for $T=2q>0$ and $b_1=-b_2$ for $T=-2q<0$, by \eqref{eq:identity_second}, the ideal coupling term is $T\bend\cdot\bsplay(\n\times\bend)$; it takes on the same value on both chiral variants and  favors both (if preceded by a \emph{positive} constant).

The elastic free-energy density that extends Frank's with the objective of describing the \NTB\ phase is thus posited as follows,
\begin{equation}
\label{eq:twist_bend_energy}
\FTB(S,T,b_1,b_2,q):=\frac12k_1S^2+\frac12k_2\Big(T^2+(2q)^2\Big)+\frac12k_3B^2+\frac14k_4\Big(T^4+(2q)^4 \Big)+\frac14k_5B^4-k_6(2q)Tb_1b_2,
\end{equation}
which, for convenience, is written in terms of the distortion characteristics.\footnote{Use of \eqref{eq:q_squared_expression} and identity \eqref{eq:identity_2} in Appendix~\ref{sec:identities} easily converts \eqref{eq:twist_bend_energy} into a formula featuring only the invariants of $(\n,\gradn)$.} The function $\FTB$ in \eqref{eq:twist_bend_energy} is deliberately built with the symmetry of the intended ground state. Thus $\FTB$ is invariant under the exchange of $T^2$ and $(2q)^2$ and the simultaneous transformations
\begin{equation}
\label{eq:twist_bend_symmetry}
(2q)T\to-(2q)T,\quad b_1b_2\to-b_1b_2.
\end{equation}
This choice makes $\FTB$ depend only on six elastic constants, only two more than in Frank's formula.\footnote{An extra quartic term, $T^2(2q)^2$, which also obeys \eqref{eq:twist_bend_symmetry}, could be added in \eqref{eq:twist_bend_energy}. But this would not alter the qualitative conclusions of the analysis that follows.} It is perhaps worth noting that unlike Frank's constants, which have physical dimensions of force, the elastic constants of the added quartic terms, that is, $k_4$, $k_5$, and $k_6$, have physical dimensions of force times length square. Thus a length scale is hidden in the theory from the start; it will reappear in the equilibrium pitch. 

A comparison of the quadratic components of \eqref{eq:twist_bend_energy} with Frank's formula \eqref{eq:Frank_energy_density_rewritten} readily identifies the constants
\begin{equation}
\label{eq:constant_identification}
k_1=K_{11}-K_{24},\quad k_2=K_{22}-K_{24}=K_{24},\quad k_3=K_{33},
\end{equation}
so that two Frank's constants should be related,
\begin{equation}
\label{eq:Frank_constants_relationship}
K_{24}=\frac12K_{22}>0.
\end{equation}
We shall also assume that $k_1>0$, so that $\FTB$ is minimized by $S=0$, as desired, and $k_2>0$, to simplify our analysis.\footnote{Letting $k_2<0$ would only prompt an annoying number of case distinctions, adding little to the variety of phases described by \eqref{eq:twist_bend_energy}.}

The leading homogeneous form in $\FTB$, the only that needs to be positive definite, is the quartic polynomial
 \begin{equation}
\label{eq:quartic_form}
\Phi:=\frac14k_4\Big(T^4+(2q)^4\Big)+\frac14k_5B^4-k_6(2q)Tb_1b_2,
\end{equation}
where we shall take $k_4$, $k_5$, and $k_6$ all \emph{positive}. As shown in Appendix~\ref{sec:quartic}, under this assumption, $\Phi$ is positive definite whenever
\begin{equation}
\label{eq:k_6_squared_inequality}
k_6^2<2k_4k_5.
\end{equation}

Let $S=0$ and set $c:=2qT$. we see that for $c=0$ $\FTB$ attains its minimum for 
\begin{equation}
\label{eq:minimizer_nematic}
T=2q=b_1=b_2=0, \quad\text{if}\quad k_3\geqq0,
\end{equation}
and for
\begin{equation}
\label{eq:minimizer_bend}
T=2q=0\quad\text{and}\quad B^2=b_1^2+b_2^2=-\frac{k_3}{k_5},\quad\text{if}\quad k_3\leqq0.
\end{equation}
The former  is the trivial uniform state, whereas the latter is a non-uniform bend state. Similarly, we see that, for given $c\neq0$, $\FTB$ attains its minimum in $(b_1,b_2)$ at $b_1=b_2=b_0$ if $c>0$ and at $b_1=-b_2=b_0$ if $c<0$, where
\begin{equation}
\label{eq:b_formula}
b_0^2=\max\left\{0,\frac{1}{2k_5}(k_6|c|-k_3) \right\}.
\end{equation}
For either sign of $c$, $\FTB$ attains the same minimum in $(b_1,b_2)$. Making use of \eqref{eq:b_formula} in \eqref{eq:twist_bend_energy}, we reduce $\FTB$ to a function $\fTB(c,q)$, even in $c$, which we need study only for $q\geqq0$,
\begin{equation}
\label{eq:f_formula}
\fTB(c,q):=\frac12k_2\left(\frac{c^2}{(2q)^2}+(2q)^2\right)+\frac14k_4\left(\frac{c^4}{(2q)^4}+(2q)^4\right)-H(k_6|c|-k_3)\frac{1}{4k_5}(k_6|c|-k_3)^2,
\end{equation}
where $H$ is Heaviside's step function.\footnote{That is, $H(x)=0$ for $x\leqq0$ and $H(x)=1$ for $x>0$.}

\begin{figure}[h]
	\centering
	\includegraphics[clip,width=.3\textwidth]{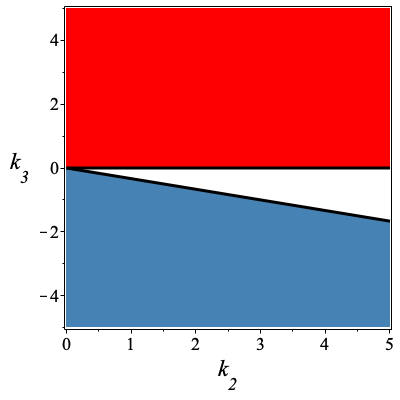}
	\caption{(Color online) Phase diagram for the minimizers of $\FTB$ in the half-plane $(k_2,k_3)$ with $k_2\geqq0$ (arbitrary units). The blue (lower) region is delimited by the straight line \eqref{eq:straight_line}. In this region, $\FTB$ is minimized by the uniform state \eqref{eq:minimizer_uniform}. In the white (middle) region, the state of minimum energy is the non-uniform pure bend \eqref{eq:minimizer_bend}. In the red (upper) region, the minimum energy is attained at the uniform state \eqref{eq:minimizer_nematic}.}
	\label{fig:phase_diagram}
\end{figure}

A simple, but tedious analysis shows that $\fTB$ attains its minimum on a uniform distortion when the elastic constants $(k_2,k_3)$ fall in two of the three regions depicted in Fig.~\ref{fig:phase_diagram}, namely, the red and blue regions.
The blue region is delimited by the straight line
\begin{equation}
\label{eq:straight_line}
k_3=-2\frac{k_5}{k_6}k_2.
\end{equation}
Below this line, $\FTB$ is minimized by
\begin{equation}
\label{eq:minimizer_uniform}
T^2=(2q_0)^2:=-\frac{k_3k_6+2k_5k_6}{2k_4k_5-k_6^2}\geqq0\quad\text{and}\quad b_1^2=b_2^2=b_0^2:=-\frac{k_2k_6+k_3k_4}{2k_4k_5-k_6^2}\geqq0
\end{equation}
In the red region the minimizer of $\FTB$ is the trivial uniform state \eqref{eq:minimizer_nematic}, while in the white region it is the non-uniform pure bend \eqref{eq:minimizer_bend}. The uniform minimizers \eqref{eq:minimizer_uniform} of $\FTB$ come in pairs, with opposite signs of $T$, confirming its double-well nature.

For given $k_2>0$, upon decreasing $k_3$ from the red region towards the blue region, as soon as we hit $k_3=0$, the ground state of $\FTB$ starts growing a preferred bend vector, whose length is prescribed according to \eqref{eq:minimizer_nematic}, while both  twist $T$ and biaxial splay $q$ remain zero, as long as we stay in the white region. Upon crossing the border of the blue region, both $T^2$ and $(2q)^2$ start growing away from zero, while keeping equal to one another. Two separate ground states develop, which have the same energy; they are characterized by the uniform heliconical fields, with \emph{different} helix axes, described in Sec.~\ref{sec:heliconical}. The bend vector, whose length grows with no discontinuity across the blue region's border, acquires, for both variants, the appropriate components $(b_0,b_0,0)$ and $(b_0,-b_0,0)$ in the distortion frame $\dframe$.

A theory based on the elastic free-energy density $\FTB$ in \eqref{eq:twist_bend_energy} would thus predict that the \NTB\ phase arises from the standard nematic phase for sufficiently negative values of $k_3$ through an intermediate non-uniform bend phase.

\section{Conclusions}\label{sec:conclusions}
It was asked in \cite{selinger:interpretation} which are all uniform nematic distortions that fill the whole space. This question was answered here by showing that the totality of such fields live in two families, each parameterized by two scalars. These fields exhaust the heliconical structures first envisaged by Meyer and recently recognized as possible ground states for twist-bend nematics. 

Taking full advantage of the symmetries enjoyed by uniform distortions, we proposed a simple elastic model whose free energy can admit as ground state either of two conjugated heliconical fields with opposite chirality, depending on the choice of two model parameters. This is our theory of generalized elasticity for nematics.

We showed that the proposed elastic free-energy density is not only minimized on uniform distortions: two regions in parameter space where it is are separated by one where it is not. In this latter, a pure bend is preferred, which cannot fill space uniformly, and so it is likely to produce elastic frustration, possibly relieved by defects.

\begin{acknowledgments}
It is for me a pleasure to acknowledge an enlightening discussion with Jonathan V. Selinger, who was kind enough to present his inspiring paper \cite{selinger:interpretation} to me prior to its publication. I am also indebted to Andrea Pedrini for a critical, suggestive reading of \cite{selinger:interpretation}.
\end{acknowledgments}


\appendix

\section{Construction of Uniform Distortions}\label{sec:construction}
Here we provide the details needed to construct the heliconical nematic field that corresponds to a given uniform distortion. In essence, we integrate \eqref{eq:decomposition_gradient_reproduced} in a \emph{fixed} frame, once the distortion characteristics have been chosen according to \eqref{eq:families}.

Our analysis builds heavily on the properties of the tensor $\Spin$ in \eqref{eq:Omega_formula}, especially on its having a set of eigenvectors that span the whole space. Let $(\lambda,\e)$ be one eigenpair of $\Spin$. Along the ray 
\begin{equation}\label{eq:ray}
p(s)=o+s\e,
\end{equation}
which passes through the origin $o$ for $s=0$,
\begin{equation}\label{eq:n_along_ray}
\n'(s)=\lambda\e\times\n(s),
\end{equation}
where a prime $'$ denotes differentiation with respect to $s$. As shown in Sec.~\ref{sec:heliconical}, along the ray \eqref{eq:ray}, the whole distortion frame precesses at the rate $\lambda$ around $\e$, so that the latter keeps constant components in that specific mobile frame (as it does in all fixed frames).

Choosing a fixed Cartesian frame $(\e_x,\e_y,\e_z)$ so that it coincides with the distortion frame at $s=0$, we thus obtain that 
\begin{equation}
\label{eq:n_delivered}
\n(s)=\R(\lambda s)\e_z,
\end{equation}
where $\R(\lambda s)$ is the rotation of angle $\lambda s$ about $\e$. This rotation can explicitly be represented as (see, for example, \cite[p.\,95]{virga:variational})
\begin{equation}
\label{eq:rotation_representation}
\R(\lambda s)=\tensor{I}+\sin(\lambda s)\W(\e)\e_z+(1-\cos(\lambda s))\W(\e)^2\e_z,
\end{equation}
where $\tensor{I}$ is the identity and $\W(\e)$ is the skew-symmetric tensor associated with $\e$. Since the components $(e_1,e_2,e_3)$ of $\e$ in the mobile distortion frame $\dframe$ as obtained in Sec.~\ref{sec:heliconical} (for the different families of uniform distortions) are the same as the components in the fixed frame $(\e_x,\e_y,\e_z)$, we can represent $\W(\e)$ as
\begin{equation}
\label{eq:W_representation}
\W(\e)=-e_1(\e_y\otimes\e_z-\e_z\otimes\e_y)+e_2(\e_x\otimes\e_z-\e_z\otimes\e_x)-e_3(\e_x\otimes\e_y-\e_y\otimes\e_x).
\end{equation}
Combining \eqref{eq:n_delivered}, \eqref{eq:rotation_representation}, and \eqref{eq:W_representation}, we readily arrive at
\begin{equation}
\label{eq:n_construction}
\n(s)=[e_2\sin(\lambda s)+e_3e_1(1-\cos(\lambda s))]\e_x-[e_1\sin(\lambda s)-e_2e_3(1-\cos(\lambda s))]\e_y+[\cos(\lambda s)+e_3^2(1-\cos(\lambda s))]\e_z.
\end{equation}
It should be noted that $\n$, as delivered by \eqref{eq:n_construction}, is invariant under the simultaneous reversion of $s$ and $\e$.

The illustrations in Figs.~\ref{fig:heliconicals} and \ref{fig:heliconical_b_0} in Sec.~\ref{sec:heliconical} have been obtained by applying formula \eqref{eq:n_construction} to the relevant eigenpair $(\lambda,\e)$ of $\Spin$.

\section{Three Identities}\label{sec:identities}
This appendix is devoted to the proof of three identities involving the distortion characteristics. Two of these identities are cubic in those characteristics, whereas the third is sextic. The first two have indeed been used in the main body of this paper, whereas the third has not. All three identities are considered together because their proof is very similar. 

We recall two classical identities valid for any smooth unit vector field $\n$ (see, for example \cite[p.\,115]{virga:variational}),
\begin{subequations}\label{eq:identities_classical}
\begin{gather}
(\gradn)\n=-\n\times\curl\n=-\bend,\label{eq:identities_classica_1}\\
|\curl\n|^2=(\n\cdot\curl\n)^2+|\n\times\curl\n|^2=T^2+B^2,\label{eq:identities_classical_2}
\end{gather}
\end{subequations}
where, representing $\bend$ as in \eqref{eq:b_components}, we have set \begin{equation}\label{eq:B_definition}
B^2=\bend\cdot\bend=b_1^2+b_2^2.
\end{equation}
Moreover, from \eqref{eq:b_components} and the definition of $T$, we obtain two equivalent expressions for $\n\times\bend$:
\begin{equation}\label{eq:n_times_b_expressions}
\begin{split}
\n\times\bend=b_1\n_2-b_2\n_1&=\n\times(\n\times\curl\n)\\
&=T\n-\curl\n,
\end{split}
\end{equation}
which, in particular, entails that
\begin{equation}\label{eq:n_times_b_2}
|\n\times\bend|^2=|\curl\n|^2-T^2.
\end{equation}

Our starting point here is again the decomposition of $\gradn$ in \eqref{eq:decomposition_gradient_reproduced}. Since, $\bend\cdot\n=0$, it readily follows from  \eqref{eq:decomposition_gradient_reproduced} and \eqref{eq:b_components} that
\begin{equation}\label{eq:grad_n_b}
(\gradn)\bend=\frac12T\n\times\bend+\frac12S\bend+q(b_1\n_1-b_2\n_2).
\end{equation}
Taking the inner product of both sides of the latter equation with $\bend$, we obtain the first identity,
\begin{equation}\label{eq:identity_1}
q(b_1^2-b_2^2)=\bend\cdot(\gradn)\bend -\frac{1}{2}SB^2.
\end{equation}
Taking the inner product of both sides of \eqref{eq:grad_n_b} with $\n\times\bend$ and making use of both \eqref{eq:n_times_b_expressions} and \eqref{eq:n_times_b_2}, we obtain the second identity,
\begin{equation}
\label{eq:identity_2}
\begin{split}
2qb_1b_2&=-\n\times\bend\cdot(\gradn)\bend+\frac12TB^2\\
&=\curl\n\cdot(\gradn)\bend+\frac12TB^2,
\end{split}
\end{equation}
whose second form follows from \eqref{eq:n_times_b_2} and the identity $(\gradn)\trans\n=\zero$.

Our last identity is a consequence of a trivial algebraic fact,
\begin{equation}
\label{eq:identity_trivial}
(b_1^2-b_2^2)^2=B^4-4b_1^2b_2^2.
\end{equation}
Multiplying both sides of \eqref{eq:identity_trivial} times $q^2$ and making use of both \eqref{eq:identity_1} and \eqref{eq:identity_2}, we arrive at
\begin{equation}
\label{eq:identity_3}
q^2B^4=\left(\bend\cdot(\gradn)\bend-\frac12SB^2\right)^2+\left(\n\times\bend\cdot(\gradn)\bend-\frac12TB^2\right)^2,
\end{equation}
which, letting $\bendhat:=\bend/B$ be the unit vector of $\bend$, can also be rewritten as
\begin{equation}
\label{eq:identity_3_rewritten}
q^2=\left(\bendhat\cdot(\gradn)\bendhat-\frac12S\right)^2+\left(\n\times\bendhat\cdot(\gradn)\bendhat-\frac12T\right)^2,
\end{equation}
which expresses $q$ in terms of  invariants derived only from $\n$ and $\gradn$. We made no use of either \eqref{eq:identity_3} or \eqref{eq:identity_3_rewritten} in the main text; I record them here because they could be of future use.

In principle, once $q>0$ is obtained from \eqref{eq:identity_3_rewritten}, equations \eqref{eq:identity_1} and \eqref{eq:identity_2} could be given the compact form,
\begin{equation}
\label{eq:identities_compact}
\begin{split}
b_1^2-b_2^2&=\gamma,\\
b_1b_2=&\beta,
\end{split}
\end{equation}
where $\gamma$ and $\beta$ are assigned scalars. In the plane $(b_1,b_2)$, equations \eqref{eq:identities_compact} represent two hyperbolas, whose intersections with the circle represented by \eqref{eq:B_definition} determine both $b_1$ and $b_2$, to within a simultaneous change of sign. That the pair $(b_1,b_2)$ can only be determined intrinsically to within a sign also follows  from \eqref{eq:b_components}, as reversing the sign of both $\n_1$ and $\n_2$ does affect neither the definition of $\bsplay$ nor the orientation of the distortion frame, expressed by $\n_1\times\n_2\cdot\n$.

\section{Quartic Potential}\label{sec:quartic}
In this appendix, we determine the condition under which the quartic form 
\begin{equation}
\label{eq:quartic_form_reproduced}
\Phi=\frac14k_4T^4+\frac14k_4'(2q)^4+\frac14k_5B^4-k_6(2q)Tb_1b_2,
\end{equation}
which includes \eqref{eq:quartic_form} as a special case, is positive definite. To address this issue, we digress slightly and recall the definition of nonlinear eigenvectors and eigenvalues for a \emph{fully} symmetric tensor $\oct$ of rank $m$ over $\complex^n$.

Let $A_{i_1\dots i_m}$ be the components of $\oct$ in an orthogonal frame $(\e_1,\dots,\e_n)$. Following a rich, albeit quite recent literature \cite{cartwright:number,ni:degree,qi:eigenvalues,qi:rank,qi:eigenvalues_invariants}, which has also been summarized in a book \cite{qi:tensor}, we say that a vector $\vv$, with components $v_i$, is an eigenvector of $\oct$ if there is a $\lambda\in\complex$ such that
\begin{equation}
\label{eq:eigenvector_definition}
A_{i_1\dots i_m}v_{i_1}\dots v_{i_{m-1}}=\lambda v_{i_m},
\end{equation}
where it is understood that repeated indices are summed. If we normalize the eigenvectors of $\oct$ so that they have unit length, it is easily seen that for every eigenpair $(\lambda,\vv)$ there is an equivalent eigenpair $(t^{m-2}\lambda,t\vv)$, for any $t\in\complex$ with $|t|=1$. Over $\real^n$, the only choices for $t$ are $t=\pm1$, and only two equivalent eigenpairs are possible, with equal or opposite eigenvalues, depending on whether $m$ is even or odd, respectively. 

It was shown in \cite{cartwright:number} that if a tensor $\oct$ of rank $m\geqq3$ over $\complex^n$ has a finite number of equivalence classes of eigenpairs, their number (counted with algebraic multiplicity) is
\begin{equation}
\label{eq:eigenpair_number}
E(m,n)=\frac{(m-1)^n-1}{m-2}.
\end{equation}
For the case that interests us here, $E(4,4)=40$. Thus, a real tensor $\oct$ of rank $4$ over $\real^4$  will \emph{at most} have  $80$ eigenvectors, if they are finite, as there is no guarantee that all eigenvectors are real. More details about eigenvectors and eigenvalues of higher-rank tensors can be found in \cite{gaeta:symmetries} and \cite{walcher:eigenvectors}.

As shown in \cite{gaeta:symmetries}, the eigenvectors and eigenvalues of $\oct$ over $\real^n$ can be identified with the critical points of a homogeneous potential,
\begin{equation}
\label{eq:potential_Phi_A}
\Phi_{\oct}(\x):=A_{i_1\dots i_m}x_{i_1}\dots x_{i_m},
\end{equation}
constrained over the unit sphere $\sphere^{n-1}$. Finding the critical points of
\begin{equation}
\label{eq:potential_Psi_A}
\Psi_{\oct}(\x):=\Phi_{\oct}(\x)-\frac{m}{2}\lambda\x\cdot\x
\end{equation}
in the whole of $\real^n$ amounts to find the real eigenvectors of $\oct$. The corresponding eigenvalues are precisely the values of the Lagrange multiplier $\lambda$ needed to obey the constraint $\x\cdot\x=1$. These latter values are, as is easily seen, the values that $\Phi_{\oct}$ takes on its constrained critical points.

Now, it is easy to connect the general theory of eigenvectors and eigenvalues for higher-rank tensors with our search for a condition of positivity for $\Phi$ in \eqref{eq:quartic_form_reproduced}. This latter would simply be the request that the least real  eigenvalue of a specific fully symmetric fourth-rank tensor $\oct$ be positive.\footnote{This generalizes the connection between the positivity of a quadratic form in $\real^n$ and the positivity of the least (standard) eigenvalue of a symmetric second-rank tensor.} Taking advantage of the inequalities
\begin{equation}
\label{eq:k_4_k_5_inequalities}
k_4>0,\quad k_4'>0,\quad\text{and}\quad k_5>0,
\end{equation}
assumed in the main text, we set
\begin{equation}
\label{eq:x_variables)_definition}
x_1:=T,\quad x_2:=\sqrt[4]{\frac{k_4'}{k_4}}2q,\quad x_3:=\sqrt[4]{\frac{k_5}{k_4}}b_1,\quad x_4:=\sqrt[4]{\frac{k_5}{k_4}}b_2,
\end{equation}
so that $\Phi$ in \eqref{eq:quartic_form_reproduced} reduces to $\Phi=\frac14k_4\Phi_{\oct}$, with
\begin{equation}\label{eq:potential_Phi_A_kappa}
\Phi_{\oct}(\x)=x_1^4+x_2^4+(x_3^2+x_4^2)^2-\kappa x_1x_2x_3x_4,
\end{equation}
where we have set
\begin{equation}\label{eq:kappa_definition}
\kappa:=4\frac{k_6}{\sqrt{k_4k_5}}\sqrt[4]{\frac{k_4}{k_4'}}.
\end{equation}

The equilibrium equations associated with the potential $\Psi_{\oct}$ defined as in \eqref{eq:potential_Psi_A}, with $m=4$, are
\begin{subequations}\label{eq:equilibrium_equations_quartic_potential}
	\begin{eqnarray}
-\kappa x_2x_3x_4+4x_1^3-4\lambda x_1&=0,\\
	-\kappa x_1x_3x_4+4x_2^3-4\lambda x_2&=0,\\
	4(x_3^2+x_4^2)x_3-\kappa x_1x_2x_4-4\lambda x_3&=0,\\
	4(x_3^2+x_4^2)x_4-\kappa x_1x_2x_3-4\lambda x_4&=0.
	\end{eqnarray}
	The real solutions $(\lambda,\x)$ to these equations and the constraint
\begin{equation}\label{eq:constraint}
x_1^2+x_2^2+x_3^2+x_4^2=1
\end{equation}
\end{subequations}	
represent all critical values and critical points of $\Phi_{\oct}$. It would be tedious to list all of them; we just remark that equations \eqref{eq:equilibrium_equations_quartic_potential} enjoy a rotational symmetry, and so there are two conjugated orbits of critical points with
\begin{equation}
\label{eq:orbits}
x_1=x_2=0,\quad x_4=\pm\sqrt{1-x_3^2},\ -1\leqq x_3\leqq 1,
\end{equation}
and so the estimate in \eqref{eq:eigenpair_number} does not apply here. All other critical points are discrete.

\begin{figure}[h]
	\centering
	\includegraphics[clip,width=.3\textwidth]{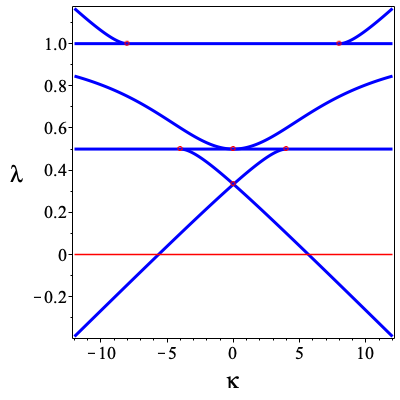}
	\caption{(Color online) The critical values of $\Phi_{\oct}$ in \eqref{eq:potential_Phi_A_kappa} as functions of the parameter $\kappa$. They are given by \eqref{eq:eigenvalues} and are all positive for $-4\sqrt{2}<\kappa<4\sqrt{2}$. All but $\lambda=1$ correspond to discrete critical points on the unit sphere \eqref{eq:constraint}. The points marked by red circles are bifurcations points, where  different eigenvalues meet and their number may change; they are located at $\kappa=0$, $\kappa=\pm4$, and $\kappa=\pm8$.}
	\label{fig:eigenvalues}
\end{figure}

Figure~\ref{fig:eigenvalues} 
represents all critical values of $\Phi_{\oct}$ in \eqref{eq:potential_Phi_A_kappa} as functions of $\kappa$. Red circles mark there the bifurcation points, which are located at $\kappa=0$, $\kappa=\pm4$, and $\kappa=\pm8$. The real eigenvalues are 
\begin{subequations}\label{eq:eigenvalues}
\begin{eqnarray}
\lambda_1&=&1,\quad\lambda_2=\frac{\kappa^2+32}{\kappa^2+64},\quad\text{and}\quad\lambda_3=\frac12,\label{eq:eigenvalues_1}\\
\lambda_4&=&\frac{32-\kappa^2}{16\kappa+96},\quad\text{for}\quad\kappa\leqq-8\quad\text{or}\quad\kappa\geqq-4,\label{eq:eigenvalues_2}\\
\lambda_5&=&\frac{\kappa^2-32}{16\kappa-96},\quad\text{for}\quad\kappa\leqq4\quad\text{or}\quad\kappa\geqq8.\label{eq:eigenvalues_3}
\end{eqnarray}
\end{subequations}
It is clear from \eqref{eq:eigenvalues_2} and  \eqref{eq:eigenvalues_3}  that all eigenvalues $\lambda$ of $\Phi_{\oct}$ in \eqref{eq:potential_Phi_A_kappa} are positive whenever $-4\sqrt{2}<\kappa<4\sqrt{2}$, or $\kappa^2<32$. Setting $k'_4=k_4$ in \eqref{eq:kappa_definition}, we thus conclude that the quadratic form $\Phi$ in \eqref{eq:quartic_form} is positive definite whenever inequality \eqref{eq:k_6_squared_inequality}  is satisfied.


%

\end{document}